# Buffer Layer Engineering on Graphene via Various Oxidation Methods for Atomic Layer Deposition

Nobuaki Takahashi[1], and Kosuke Nagashio[*1,2]

[1]*Department of Materials Engineering, The University of Tokyo, Tokyo 113-8656, Japan*
[2]*PRESTO, Japan Science and Technology Agency (JST), Tokyo 113-8656, Japan*
[*]nagashio@material.t.u-tokyo.ac.jp

**Abstract:** The integration of high-$k$ oxide on graphene using atomic layer deposition requires an electrically reliable buffer layer. In this study, Y was selected as the buffer layer due to the highest oxidation ability in rare earth elements and various oxidation methods (atmospheric, high-pressure $O_2$ and ozone) were applied to the Y metal buffer layer. By optimizing oxidation conditions of the top gate insulator, we successfully improve the capacitance of top gate $Y_2O_3$ insulator and demonstrate a large $I_{on}/I_{off}$ ratio for bilayer graphene under an external electric field.

An electrically reliable gate stack formation is a building block for graphene field-effect transistors (FETs). Recently, there are two research directions. The first is the heterostructure formation with a layered $h$-BN top-gate insulator primarily for the condensed matter physics study at low temperature under high magnetic field.[1-3] The second is the high-$k$ oxide formation on graphene for technological applications.[4-8] Thus far, we achieved the high insulating properties of high-$k$ $Y_2O_3$ top-gate in graphene FETs by the Y metal deposition and subsequent high-pressure $O_2$ post-deposition annealing (HP-PDA).[9] In addition, we succeeded in quantitatively evaluating the gap state density in the band gap of bilayer graphene.[10] However, the surface roughness of $Y_2O_3$ insulator is not sufficiently small to further reduce the effective oxide thickness (EOT) due to the high deposition rate. To control the electrical quality of high-$k$ insulator more accurately, atomic layer deposition (ALD) is one of the most widely used deposition methods because it results in excellent conformality and film-thickness controllability.[11-13] ALD suffered from nucleation difficulties because graphene surface is chemically inert.[14,15] To overcome these issues, the buffer layer has been intensively studied such as thin oxidized metal layers[12,15-18] and organic polymers.[11,14,19] Moreover, a good wetting property of $Y_2O_3$ on graphene has been recently reported.[20-22]

Although the degree of oxidation of buffer layers is critical, various oxidation methods [e.g., atmospheric-pressure $O_2$ annealing (AP), high-pressure $O_2$ annealing (HP) and ozone annealing] have not been studied yet. Based on the thermodynamic consideration, as shown in Supplementary **Fig. S1**, the oxidizing ability follows the order of ozone, HP and AP. In this study, we applied different oxidation methods to the Y metal buffer layer and fabricated the $Y_2O_3$ top-gate graphene FETs using ALD. We discuss the suitable deposition and oxidation methods for the buffer layer. Additionally, we discuss the limitation of direct deposition of high-$k$ on graphene based on the estimated $k$ value.

The ~90 nm $SiO_2$/n$^+$-Si substrates were heated at 1050 °C for 5 min in $O_2$ gas flow to obtain a hydrophobic $SiO_2$ surface with siloxane groups.[23] Then, monolayer graphene was transferred on it by mechanically exfoliating Kish graphite. We fabricated bilayer graphene back-gate FETs containing source and drain electrodes [Ni(~10 nm)/Au(~50 nm)] using conventional electron beam (EB) lithography.[24] The device was annealed in the Ar/$H_2$ mixture gas flow at 300 °C for 3 hours in the deposition chamber. Subsequently, the Y metal was deposited via thermal evaporation of the Y metal in a PBN crucible in an Ar atmosphere with a partial pressure of $10^{-1}$ Pa. The thickness was controlled to be ~1 nm at a rate of ~0.1 A/s. Then, the Y metal was oxidized via three different oxidation methods, (i) AP annealing in 100 % $O_2$ at 300 °C for 10 min, (ii) HP annealing at 100 atm in 100 % $O_2$ at 300 °C for 10 min, and (iii) Ozone annealing with ~100,000 ppm at RT for 5 min. In this study, ozone was generated using a silent discharge method (i.e., no UV irradiation). These conditions were selected to prevent the defect formation in graphene, which was confirmed by Raman measurement through the $Y_2O_3$ layer. For ozone annealing, the ozone generally becomes inactive rapidly. So, the generated ozone concentration is considerably reduced through the home-made gas line from ozone generator to the rapid thermal annealing system (RTA). After ozone annealing at 300 °C, graphene severely damaged. With reducing the temperature, $I_D/I_G$ is drastically reduced. At room temperature, D band was negligible.

Then, the sample was moved to an ALD reactor for $Y_2O_3$ deposition. (i-PrCp)$_3$Y and water were used as precursor and oxidant, respectively. The thickness of $Y_2O_3$ deposited using ALD was ~5 nm and was estimated from the predetermined relation between the $Y_2O_3$ thickness on $SiO_2$/Si wafer and deposition cycles. The detailed ALD conditions are shown elsewhere.[18] After the $Y_2O_3$ deposition, the Al top gate electrode was formed via the EB lithography. Then, the measurement was performed in a vacuum prober at RT.

First, the suitable buffer layer deposition methods were explored to prevent the defect formation in graphene. Monolayer graphene is used in this study because it is most sensitive to defect formation. The electron beam evaporation (EB) of the Y metal with ~0.1 A/s at $10^{-5}$ Pa on monolayer graphene introduced defects ($I_D/I_G$=0.355), as evident from the Raman D band in **Fig. 1(a)**. It has been reported that the grazing-angle radio frequency sputtering strongly reduces the defect formation due to the angle dependence of energies required to displace a single C atom out of graphene.[25] Therefore, the sample was mounted with tilt angles of 45°, 60° and 90°, and then the Y metal was deposited again. With increasing tilt angles, the D band intensity was reduced, but the defects were still formed at 60° ($I_D/I_G$=0.029). No Y deposition was observed at 90°. Although EB evaporation is widely used for the buffer layer deposition of metals with high melting points, the defect formation should be taken into account.

In case of EB evaporation, X-ray produced during the EB evaporation and/or Y vapor with the high kinetic energy might be origins for the defect formation. Therefore, simple thermal evaporation with a PBN crucible was performed in vacuum ($10^{-5}$ Pa), as shown in **Fig. 1(b)**. The D band was still observed. These results suggest that the kinetic energy of Y vapor should be reduced. Therefore, Ar was introduced into the chamber, and the partial pressure of Ar was kept as $10^{-1}$ Pa. Thus, the mean free path was approximately 30 mm. In this case, no D band was found in the Raman data. In case of the film deposited by Ar sputtering, the capture of Ar in the film often reduce its electrical quality. The capture of Ar in the Y metal film was analyzed using thermal desorption spectroscopy and no Ar was detected. **Figure 1(c)** shows the atomic force microscopy (AFM) image of 1 nm Y metal on graphene, where the surface roughness (RMS) is ~0.3 nm. Based on these results, the thermal deposition of the Y metal in Ar is suitable for the buffer layer deposition. It should be noted that the thermal evaporation of the Y metal in $O_2$ (partial pressure of $10^{-1}$ Pa) is not possible because the low deposition rate cannot be maintained due to the rapid oxidation of Y metal in a PBN crucible.

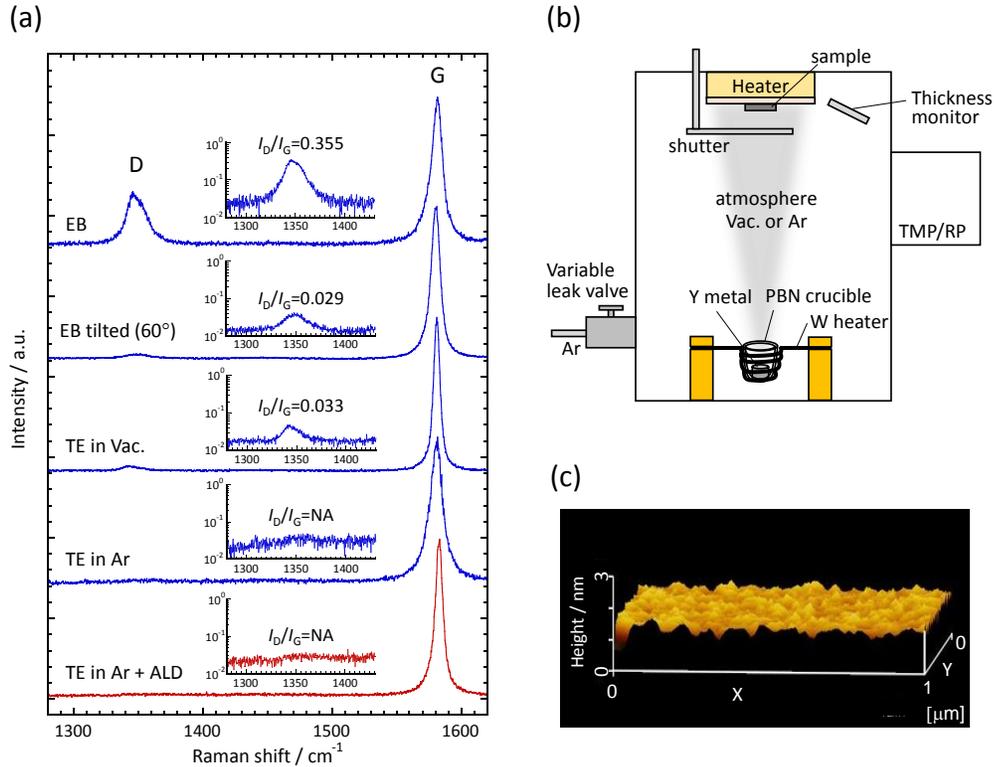

**Figure 1**. (a) Raman data for monolayer graphene with buffer layers deposited at different conditions. "EB" and "TE" indicate electron beam evaporation and thermal evaporation, respectively. The inset shows the magnified figure around the D band (vertical axis is in the log scale). The bottom data are obtained after the thermal deposition of the Y buffer layer in Ar, AP annealing, and the $Y_2O_3$ deposition using ALD. (b) Schematic drawing of the Y metal deposition chamber. (c) AFM image of the surface of the Y metal buffer layer deposited in Ar at a partial pressure of $10^{-1}$ Pa.

Next, the oxidation of the Y metal buffer layer on monolayer graphene was investigated. The oxidation ability of Y metal is largest in the typical rare earth metals. Therefore, the Y metal is generally oxidized after the sample is taken out from the chamber, maybe even in the chamber with the vacuum of $10^{-5}$ Pa. Here, in order to improve the oxidation condition further, three different oxidation conditions were selected as mentioned above.

To reveal the effect of oxidation conditions of the buffer layer on the electrical quality of top gate insulator, the top-gate $Y_2O_3$ insulator with ~5 nm thickness was deposited using ALD. No Raman D band was observed after the oxidation of buffer layer and subsequent $Y_2O_3$ deposition using ALD, as shown in **Fig. 1(a)**. It should be noted that all devices were fabricated for the bilayer graphene because bandgap opening is our research motivation. The inset in **Figure 2(a)** is a scanning electron microscope (SEM) image of the typical dual gate bilayer graphene FET. The channel length and width for this device is 10 μm and 2 μm, respectively. The typical dimension in this study is ~10 mm × 5 mm. **Figure 2(a)** shows the resistivity as a function of top gate voltage ($V_{TG}$) for the device in which the buffer layer was oxidized via

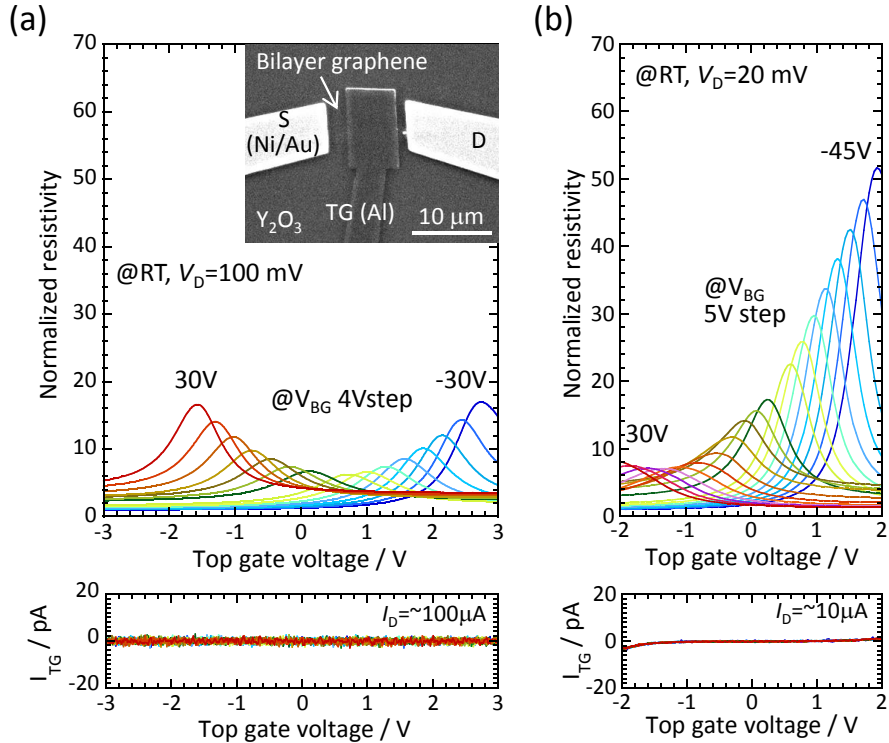

**Figure 2**. (a) Normalized resistivity and top-gate leakage current as a function of $V_{TG}$ for different $V_{BG}$ at RT ($V_D$ = 100 mV). The top-gate insulator formation condition is AP annealing of Y buffer layer and subsequent ALD-$Y_2O_3$. It should be noted that resistivity is normalized by the value of the lowest resistivity value at $V_{TG}$ = -3 V and $V_{BG}$ = -30 V. The voltage step for $V_{BG}$ is 4 V. The inset SEM image is the fabricated top-gate device. (b) Normalized resistivity and top-gate leakage current as a function of $V_{TG}$ for different $V_{BG}$ at RT ($V_D$ = 20 mV). The top-gate insulator formation condition is HP annealing of both the buffer layer and ALD-$Y_2O_3$. The voltage step for $V_{BG}$ is 5 V.

AP annealing. The resistivity increase at the Dirac point is clearly observed with changing the back gate voltage ($V_{BG}$). The leakage current through the top gate ($I_{TG}$) is negligible, which suggests that an electrically reliable top-gate insulator is formed. For the other two oxidation conditions (HP annealing and ozone annealing), similar transfer characteristics were observed.

**Figure 3(a)** shows the trace of the Dirac point observed for the $V_{TG}$ sweep at different $V_{BG}$ for different oxidation conditions (AP, HP and ozone annealing). The Dirac point position is controlled by the relative ratio of capacitive coupling between the top and back gates with bilayer graphene. Therefore, the slope ($S$) in **Fig. 3(a)** corresponds to - $C_{BG}/C_{TG}$. $C_{BG}$ is calculated to be 0.0384 μF/cm$^2$ for the 88 nm thick SiO$_2$ using $k_{SiO2}$ = 3.9. $C_{TG}$ values that are estimated for different oxidation conditions are shown in **Fig. 3(b)**. For the transport measurements on AP annealed devices, the almost whole percentage of devices works well, while the capacitance value was small. The thickness of ALD-Y$_2$O$_3$ are reduced from 6 nm to 5 nm for the HP and ozone annealing cases. The estimated $k$ values are shown in **Fig. 3(b)**. In this study, the Y$_2$O$_3$ thickness on graphene could not be determined directly due to the small area of graphene channel. Therefore, the error bar indicates ±15 % of the value, which approximately corresponds with the error of ±1 nm in thickness. The values of $k$ for HP and ozone were higher than those of AP. However, $k$ for ozone with the highest oxygen potential is lower than that for HP. In general, the reactivity of ozone drastically reduced with temperature, as mentioned above. Therefore, we expect that ozone does not reach to the bottom of the 1-nm thick Y metal buffer layer. Therefore, ozone annealing at RT was not effective from the viewpoint of oxidation kinetics, although oxygen potential is the highest for ozone annealing under three oxidation conditions. In the previous study, we achieved the $k$ value of 12.9 for the Y$_2$O$_3$ deposition via ALD on SiO$_2$/Si substrate without the post-deposition annealing.[18] This value is close to the ideal value for the bulk Y$_2$O$_3$ and is much higher than those in the present study. This suggests that the quality of ALD-Y$_2$O$_3$ is strongly affected by the buffer layer quality. Considering that Y has the highest oxidation ability in rare earth elements, and Y$_2$O$_3$ is thermodynamically stable on graphene,[9] it seems to be difficult to achieve the full buffer layer oxidation without introducing defects in graphene.

To further improve the electrical quality of top gate insulator, the 5-nm Y$_2$O$_3$ film was deposited using ALD after the Y buffer layer was oxidized in AP annealing. Then, HP annealing was performed for the entire top-gate insulator (the buffer layer and ALD-Y$_2$O$_3$). It is noted that the annealing conditions for AP and HP are the same as mentioned above. As shown in **Fig. 2(b)**, the modulation of resistivity is clearly improved due to the high top-gate capacitance, compared with that in **Fig. 2(a)**. The maximum $I_{on}/I_{off}$ ratio is ~50. The negligible $I_{TG}$ suggests that the electrically reliable top-gate insulator is formed. Among the several devices, the best value for $C_{TG}$ was 1.16 μF cm$^{-2}$ ($k$ = ~7.8), as shown in **Fig. 3(b)**. The value of $k$ for the whole HP annealing for both the buffer layer and ALD-$Y_2O_3$ was improved compared with $k$ for HP annealing only for the buffer layer. This suggests that the degree of oxidation of the buffer layer largely influences the electrical quality of ALD-$Y_2O_3$ as well.

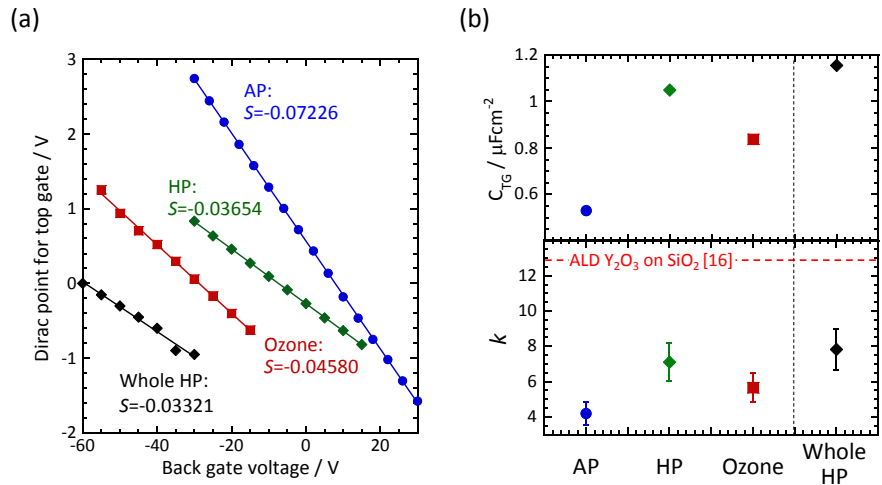

**Figure 3**. (a) Trace of the Dirac point observed for the $V_{TG}$ sweeps at different $V_{BG}$ for different oxidation conditions (AP, HP ozone annealing, and whole HP). The slopes indicated in this figure are determined via linear fitting. (b) $C_{TG}$ (top) and $k$ (bottom) estimated for different oxidation conditions. It should be noted that the thickness of buffer layer and ALD-Y$_2$O$_3$ are 1 nm and 6 nm for the AP annealing case, while they are 1 nm and 5 nm for the HP and ozone annealing cases.

The high insulating properties of high-$k$ $Y_2O_3$ top-gate in graphene FETs was previously achieved by the Y metal deposition and subsequent HP-PDA without using ALD.[9] Now let's compare the present whole HP annealed ALD device with the previous HP-PDA device. As shown in **Figure 4(a)**, the initial leakage current level is the same for both cases. However, at the high electrical field, the leakage current for the whole HP annealed ALD device increases more rapidly. The breakdown voltage for the whole HP annealed ALD device is lower than that for the HP-PDA device. Although the present ALD provides the much precise thickness control, the previous HP-PDA device is preferable in terms of electrical quality. This can be understood by the cross-sectional transmission electron microscope (TEM) image of the whole HP annealed ALD device, as shown in **Fig. 4(b)**. The total thickness of $Y_2O_3$ layer was consistent with our expectation (~6 nm). The wettability of Y metal on graphene is good, as shown in **Fig. 1(c),** which is consistent with previous reports.[20-22] Therefore, the surface condition for the subsequent ALD deposition is expected to be reasonable, resulting in the precise thickness control. On the other hand, it is evident that the whole $Y_2O_3$ insulator is still mainly amorphous for the whole HP annealed ALD device, while the HP-PDA device showed high density of crystals (cubic phase).[9] This is a little surprising because the partial crystallization for ALD-$Al_2O_3$ without post-annealing has been reported.[17] In general, the amorphous phase is required as a top gate insulator because the leakage current through the grain boundaries are minimized. However, the electrical quality of the present amorphous $Y_2O_3$ on graphene is rather poor. So, the crystalline $Y_2O_3$ in the HP-PDA device is still better than the amorphous $Y_2O_3$ in the whole HP annealed ALD device. Further study to improve the electrical quality of the amorphous high-$k$ on graphene is required from the chemical viewpoint.

The comparison of $C_{TG}$ with the previously reported values for monolayer, bilayer and trilayer graphene is shown in Supplementary **Fig. S2**. All the reported data do not reach the effective oxide thickness (EOT) of 1 nm, which is a standard value for Si FETs. Although there are two reports with $C_{ox} > 2$ $\mu Fcm^{-2}$, one case shows the increase in the D band intensity after the ALD process[26] and the other case uses the EB evaporation to deposit the top gate insulator.[27] Therefore, under the restraint condition for no defect formation in graphene, $C_{TG}$ obtained in this study is relatively high compared with previous reports in which $HfO_2$ with much higher $k$ value was used. As mentioned in the *Introduction*, there are two directions for the gate stack research: high-$k$ ALD oxide and $h$-BN. However, $C_{TG}$ for $h$-BN is generally small due to the small $k$ value (3~4). To overcome the oxidation issue for the buffer layer, the combination of $h$-BN and high-$k$ oxide is the key because the oxidation barrier of $h$-BN is high (> 800 °C).[28,29] Although high-$k$ oxide deposition on $h$-BN using ALD has been achieved,[18,30] $C_{TG}$ for high-k on $h$-BN is still low due to the thick *h*-BN, as shown in the Supplementary **Fig. S2**.[31,32] As we have already proposed,[18] the monolayer *h*-BN is critical for this issue.

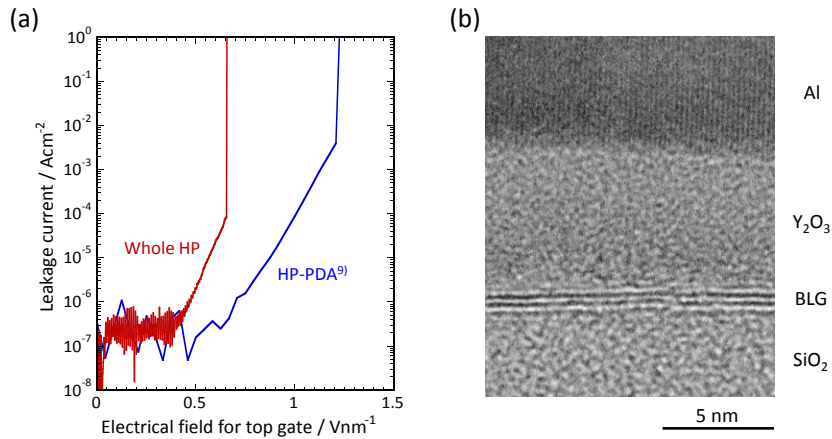

**Figure 4**. (a) Leakage current density through the $Y_2O_3$ topgate insulator for the whole HP annealed ALD device (HP annealing for buffer layer and ALD-$Y_2O_3$). For comparison, the data of the previous HP-PDA device is also shown.[9] (b) Cross-sectional TEM image for whole HP annealed ALD device.

Based on many types of trials for the buffer layer deposition, the thermal deposition of rare earth Y metal at the partial pressure of $10^{-1}$ Pa for Ar was found to be optimum because defect formation in graphene can be avoided by reducing the kinetic energy of Y vapor. Various oxidation methods were applied to the Y metal buffer layer and, then, the top gate insulator was deposited using ALD. The estimated dielectric constants for the $Y_2O_3$ films deposited on buffer layers via ALD strongly depend on the degree of oxidation of the buffer layers. By optimizing oxidation conditions of the top gate insulator, we demonstrate a large $I_{on}/I_{off}$ ratio for bilayer graphene by applying the external electric field. However, because the buffer layer is oxidized under the restraint condition for no defect formation in graphene, it is difficult to achieve an ideal dielectric constant for $Y_2O_3$. To further improve the degree of oxidation for the buffer layer, the combination of *h*-BN and high-*k* oxide is key because the oxidation barrier of *h*-BN is high.

**Acknowledgements**
We are grateful to Covalent Materials for kindly providing us Kish graphite. This research was partly supported by JSPS KAKENHI Grant Numbers JP25107004, JP16H04343, & JP16K14446 and also by JSPS Core-to-Core Program, A. Advanced Research Networks.

# Buffer Layer Engineering on Graphene via Various Oxidation Methods for Atomic Layer Deposition

Nobuaki Takahashi[1], and Kosuke Nagashio[*1,2]

[1]*Department of Materials Engineering, The University of Tokyo, Tokyo 113-8656, Japan*
[2]*PRESTO, Japan Science and Technology Agency (JST), Tokyo 113-8656, Japan*
[*]nagashio@material.t.u-tokyo.ac.jp

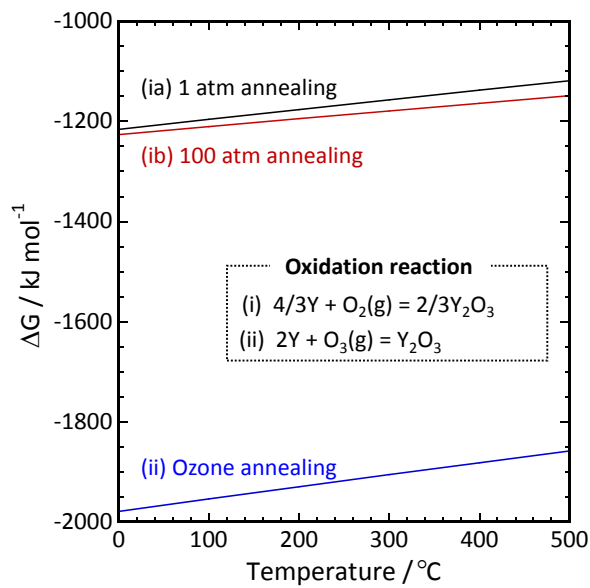

**Figure S1**. (a) $\Delta G$ of three different oxidation conditions (ia) AP ($O_2$: 1 atm), (ib) HP ($O_2$: 100 atm), and (ii) ozone ($O_3$: 100,000 ppm) at different temperatures calculated using a thermodynamic database [S1-S3]. The considered oxidation reaction equations are shown in the figure. $\Delta G$ should be a negatively larger value to facilitate oxidation. The oxidizing ability follows the order of ozone, HP and AP.

Supplementary information

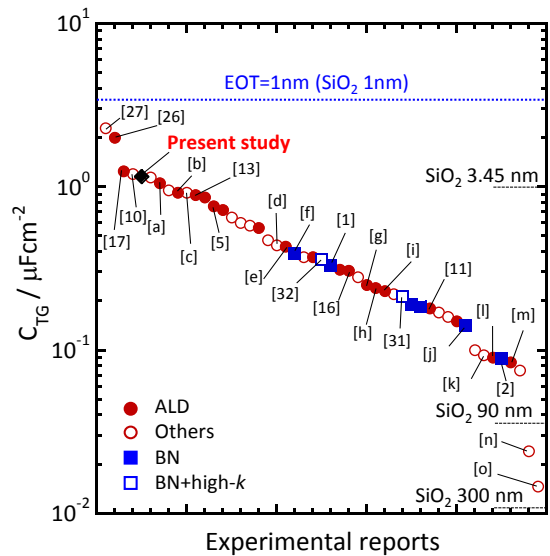

**Figure S2**. Comparison of $C_{TG}$ with the previously reported values for monolayer, bilayer and trilayer graphene. Closed and open red circles indicate $C_{TG}$ obtained for oxide insulators deposited via ALD and for insulators prepared using another technique, respectively. Closed and open blue boxes indicate $C_{TG}$ obtained for *h*-BN and for the combination of *h*-BN and high-*k* oxide. It should be noted that ref. 30 is not obtained for graphene but for $MoS_2$. "EOT = 1 nm" indicates $C_{TG}$ obtained for 1 nm thick $SiO_2$. The reference numbers in **Fig. S2** are consistent with the reference numbers in main text, while the other alphabetical references are shown below.

a) T. Chu, and Z. Chen, Nano Res. **8**, 3228 (2015).
b) S. Jandhyala, G. Mordi, B. Lee, G. Lee, C. Floresca, P. -R. Cha, J. Ahn, R. M. Wallace, Y. J. Chabal, M. J. Kim, L. Colombo, K. Cho, and J. Kim, ACS nano **6**, 2722 (2012).
c) S. L. Li, H. Miyazaki, M. V. Lee, C. Liu, A. Kanda, and K, Tsukagoshi, Small **7**, 1552 (2011).
d) L. Liao, Y. -C. Lin, M. Bao, R. Cheng, J. Bai, Y. Liu, Y. Qu, K. L. Wang, Y. Huang, and X. Duan, Nature **467**, 305 (2010).
e) B. N. Szafranek, G. Fiori, D. Schall, D. Neumaier, and H. Kurz, Nano Lett. **12**, 1324 (2012).
f) H. Wang, T. Taychatanapat, A. Hsu, K. Watanabe, T. Taniguchi, P. Jarillo-Herrero, and T. Palacios, IEEE Electron Device Lett. **32**, 1209 (2011).
g) Y. -M. Lin, A. Valdes-Garcia, S. -J. Han, D. B. Farmer, I. Meric, Y. Sun, Y. Wu, C. Dimitrakopoulos, A. Grill, P. Avouris, and K. A. Jenkins, Science **332**, 1294 (2011).
h) K. Lee, B. Fallahazad, H. Min, and E. Tutuc, IEEE Trans. Electron Devices **60**, 103 (2013).
i) T. Taychatanapat, and P. Jarillo-Herrero, Phys. Rev. Lett. **105**, 166601 (2010).
j) G. L. Yu, R. Jalil , B. Belle, A. S. Mayorov, P. Blake, F. Schedin, S. V. Morozov, L. A. Ponomarenko, F. Chiappini, S. Wiedmann, U. Zeitler, M. I. Katsnelson, A. K. Geim, K. S. Novoselov, and D. C. Elias, Proc. Natl. Acad. Sci. USA **110**, 3282 (2013).
k) S. Takabayashi, S. Ogawa, Y. Takakuwa, H. -C. Kang, R. Takahashi, H. Fukidome, M. Suemitsu, and T. Otsuji, Diamond & Relat. Mater. **22**, 118 (2012).
l) Y. Wu, Y. -M. Lin, A. A. Bol, K. A. Jenkins, F. Xia, D. B. Farmer, Y. Zhu, and P. Avouris, Nature **472**, 74 (2011).
m) Y. Zhang, T. -T. Tang, C. Girit, Z. Hao, M. C. Martin, A. Zettl, M. F. Crommie, Y. R. Shen, and F. Wang, Nature **459**, 820 (2009).
n) J. Yan, and M. S. Fuhrer, Nano Lett. **10**, 4521 (2010).
o) J. Velasco Jr, L. Jing, W. Bao, Y. Lee, P. Kratz, V. Aji, M. Bockrath, C. N. Lau, C. Varma, R. Stillwell, D. Smirnov, F. Zhang, J. Jung, and A. H. MacDonald, Nat. Nanotechnol. **7**, 156 (2012).